\documentstyle[preprint,aps]{revtex}
\begin{document}
\draft
%\parskip=4pt
%\parindent=18pt
%\baselineskip=21pt
%\setcounter{page}{1}
%\vspace{4ex}
%\vspace{20mm}
%\bigskip
\title{A mean field theory for the spin ladder system
}
%\vspace{10mm}
\author{Xi Dai and  Zhao-bin Su}
%\vspace{5mm}
\address{
 Institute of Theoretical Physics, Academia Sinica,
P.O. Box 2735, Beijing 100080, China\\
}
\maketitle
%\vspace{4ex}
%\vspace{15mm}
%\bigskip
%\vspace{4ex}

\begin{abstract}
%\centerline{\large\bf   Abstract}

In the present paper, we propose a mean field approach for spin ladders based
upon the Jordan-Wigner transformation along an elaborately ordered path.
 We show on the mean field level
that ladders with even number legs open a energy gap in their low energy 
excitation with a magnitude close to the corresponding experimental values,
 whereas the low energy excitation of the odd-number-leg ladders
are gapless. It supports the validity of our approach. We then
 calculate the gap size and the 
excitation spectra of
2-leg-ladder system. Our result is in good agreement with both the
experimental data and the numerical results. 
\end{abstract}

\vspace*{.8cm}

{\bf Category}: Ca2

{\bf Contact Author}: Xi Dai

{\bf Mailing Address}: Institute of Theoretical Physics,
 P.O. Box 2735, Beijing 100080,\\
\hspace*{4cm}  P. R. of China

{\bf E-mail}: daix@sun.itp.ac.cn

{\bf Fax}:86-10-62562587

\newpage

\centerline{\bf I. INTRODUCTION}
\vspace{.5cm}

The study on low dimensional Heisenberg anti-ferromagnetic 
model is one of the  most active research field in condensed 
matter physics .Haldane\cite{hal} conjectured that for integer-spin
 one dimensional anti-Ferromagnetic chain a energy gap exists in the
 low energy excitation spectrum , but for half integer-spin 
case the excitation spectrum is gapless. The spin 1/2 anti-Ferromagnetic 
chain can be solved exactly by Bathe ansatz\cite{bethe}.
 The excitation spectrum
is found to be gapless. The measurement on the realistic ladder material
such as $SrCu_2O_3$(two-leg), $Sr_2Cu_3O_5$(three-leg), or$(VO)_2P_2O_7$
(two-leg)\cite{exp1} shows that the spin excitation gap is opened in the ladders with
even numbers of legs, while for ladders with odd numbers of legs, no gap is found
in spin excitation. This conclusion is predicted by the early numerical 
calculation\cite{num1} and was explained qualitatively by Khveshchenko
\cite{khv}. In Khveshchenko's explanation a topological term appears in the
effective Hamiltonian of the long wavelength dynamics in odd-leg ladder
and is absent in even-leg ladder. Recently, G.Sierra\cite{sierra} has mapped the ladder
problem onto a effective one dimensional non-linear Sigma model. An extra 
topological term appears in odd-leg system and is absent in even-ladder system.
 This difference between the odd and even-
leg ladders is essentially an extension for the difference between the half
 integer and integer 
Heisenberg chains.

We also have the experience from the 1-D quantum Heisenberg spin chain, that
it is not so trivial to incorporate the subtle physics of the topological
 term in a mean field approach.
For two-leg ladder, the existence of energy gap in spin excitation for
nonzero inter chain coupling J' is 
confirmed  by various method such as Lanczos, quantum Monte Carlo
\cite{num1,num2},renormalization group\cite{ren}, variational method
\cite{var},strong coupling expansion\cite{rei}, spin liquid mean field approach\cite{sig}, as well as the 
mean field approach based on the Jordan-Wigner transformation\cite{azz}.  
And for J=J' (J is
the exchange coupling along the chains) which is the physical parameter
of the real ladder compounds, the energy gap obtained by numerical 
calculation is 0.5J which is very close to the experimental value. The
spin wave excitation spectra has also been obtained by the numerical 
calculation
which shows the minimum of the spectra is located at the wave number $\pi$
\cite{num1,num2}.
This was also confirmed by the recent neutron scattering experiment\cite{neu}.

A mean field treatment of the two-leg Heisenberg ladder with the application 
of Jordan-Wigner(J-W) transformation has been proposed by 
M.Azzouz and et al\cite{azz}. In their approach the Jordan-Wigner
 transformation
is introduced to map the spin 1/2 system to the spinless Fermion system.
The gap is obtained in their mean field approach which is about 0.7J' under
the case of J=J' which does not fit well with the experimental value. The 
excitation spectra is also calculated in their approach, which contents
a minimum at the wave number $\pi$. But the shape of the spectra is not
consistent with the numerical result which predicts the maximum of the
spectra locating between the 0 and $\pi/2$\cite{num2}.

In the present paper, we propose a mean field approach also
 based on the J-W transformation\cite{jor}.
 Our mean field approach is quite different
from the approach used in paper\cite{azz}. When performing J-W 
transformation in the ladder system, one must put the sites in a queue. Then
the spin operator can be expressed as $S^+_i=c^+_ie^{i\pi\hat\Phi (i)}$, 
where $\hat\Phi(i)$ is
nothing but the summation of number of the sites from $-\infty$ to the (i-1)th
site in the particularly ordered queue which
 are occupied by spinless Fermions. 
The difference of particularly elaborated queues used in our approach
 and  that in paper
\cite{azz} are shown in Fig.1.(a) and (b). In our approach the
sites in the odd number rungs are labeled from upward to downward but
from downward to upward in the even number rungs(shown as Fig.1(a)),
 whereas in paper
\cite{azz} all rungs are labeled from upward to downward.(shown as Fig.2(b)).

 We perform the J-W transformation with such a specially ordered 
queue. The 
advantages of our approach are as the following. First of all if we 
replace the phase 
of the hopping term by its average value, we can easily obtain the mean field
Hamiltonian for n-leg ladders in which the effective hopping terms of the 
spinless Fermions have alternative signs along both the chain and rung 
directions. This mean field Hamiltonian is similar with the one used in 
paper\cite{azz} for 2-leg case. But in their approach a further assumption,
i.e. adding a  $\pi$-flux to each plaquette is needed to obtain such 
kind of mean field Hamiltonian, whereas
in our approach the mean field Hamiltonian is obtained directly  by replacing
the phase factor in the hopping terms with their average values. We can then
 easily show that the spin gap is only opened in even-leg ladders. In certain
degree, it supports the validity of our way of Jordan-Wigner transformation
construction. 

Secondly  we can treat the phase factor in the hopping terms more
carefully by introduce a self consistent procedure for 2-leg ladder. 
We find that both the gap
magnitude and the spin excitation spectra are in good agreement with the 
numerical and experimental results in the case of J'=J which is of
 the physical 
parameter for the real ladder compounds. Actually he energy gap is found 
to be 0.46J in the case  J'=J
which is very close to the experimental result $0.47\pm 0.2J$. 
The spin wave excitation spectra obtained by our mean field 
calculation is also consistent with the numerical result with the minimum
at the wave number $\pi$ and the maximum at the wave number $0.356\pi$.
Compared with the approach used in \cite{azz}, our mean field approach 
is much better in the case of $J'>0.5J$. In the weak coupling regime our
approach works not so good, the energy gap is even persisted in 
the case of $J'=0$.
In spite of the unsatisfactory aspect in the weak coupling regime, 
our approach
is still valuable because it works very well in the intermediate and strong
coupling regime, which has the experimental correspondence.

In section II, we calculate the spin excitation gap of the spin ladders
with various numbers of legs and show that in the mean field level   
the energy gap only exists in ladder system with even number legs.
In section III, we propose a more carefully treatment of the 2-legs
spin ladder. The spin excitation gap, as well as the excitation spectra
is calculated. Finally, we make the conclusion remarks in Sec.IV.

\newpage
\centerline{\bf II.Mean field treatment for the spin gap in n-leg ladders}

We begin with the 2M-leg anti-ferromagnetic Heisenberg ladder Hamiltonian:

$$
H=J'\sum_i\sum_{p=1}^{2M-1}\vec S_{i,p}\cdot\vec S_{i,p+1}+
  J\sum_i\sum_{p=1}^{2M}\vec S_{i,p}\cdot\vec S_{i+1,2M+1-p}
$$

$$
~=J'\sum_i\sum_{p=1}^{2M-1}S^z_{i,p}\cdot S^z_{i,p+1}+
  J\sum_i\sum_{p=1}^{2M}S^z_{i,p}\cdot S^z_{i+1,2M+1-p}+
$$

\begin{equation}
~\frac{J'}{2}\sum_i\sum_{p=1}^{2M-1}S^+_{i,p}\cdot S^-_{i,p+1}+
~\frac{J}{2}\sum_i\sum_{p=1}^{2M}S^+_{i,p}\cdot S^-_{i+1,2M+1-p}+H.C.
\end{equation}
In the above Hamiltonian, `i' represents the site position along the
chains and 'p' represents the 2M sites of different chains coupled by
the inter-chain coupling constant J'. As shown in Fig.1, p is labeled
from upward to downward at the even sites while downward to upward
at the odd sites.
This is different from what used in the paper of Azzouz et.al.\cite{azz}.   
Then we introduce the generalized J-W transformation:
\begin{equation}
S_{p,i}^+=c_{p,i}^+e^{i\pi\sum_{n=-\infty}^{i-1}\sum_{l=1}^{2M}
c_{l,n}^+c_{l,n}+i\pi\sum_{l=1}^{p-1}c_{l,i}^+c_{l,i}}
\end{equation}
in which c is the spinless Fermion operator. The summation in the phase 
factor is nothing but the number of the occupied sites before the ith site
along the particular queue shown in Fig.1(a). Then the quantum spin 1/2
Hamiltonian can be mapped onto a spinless Fermion Hamiltonian as:

$$
H=J'\sum_{i,p=1}^{2M-1}({1 \over 2} -c^+_{i,p}c_{i,p})\cdot({1 \over 2}
-c^+_{i,p+1}c_{i,p+1})+
  J\sum_{i,p=1}^{2M}({1 \over 2} -c^+_{i,p}c_{i,p})\cdot({1 \over 2}
-c^+_{i+1,2M+1-p}c_{i+1,2M+1-p})+
$$

\begin{equation}
\frac{J'}{2}\sum_{i,p=1}^{2M+1}(c_{i,p}^+c_{i,p+1}+H.C.)+\frac{J}{2}\sum
_{i,p}^{2M}(c_{i,p}^+c_{i+1,2M+1-p}e^{-i\hat\Phi(p)}+H.C.)
\end{equation}
where 
$$
\hat\Phi(p)=\pi\sum_{l=p+1}^{2m}(n_{i,l}+n_{i+1,2M+1-l})
$$

In our mean field approach, we replace $n_{i,l}$  by $<n_{i,l}>$. For the 
present study, we further assume that the finite magnetization 
in this system is not 
possible because of the strong quantum fluctuation. This is reasonable for
systems with its leg number much less then the site number of each chain.
Then we have $<S_{i,l}^z>=<1/2-c_{i,l}^+c_{i,l}>=0$ which implies 
$<n_{i,l}>$=0.5. Consequently, the phase factor in eq.(3) can be 
approximated by:$\Phi(p)=\pi(2M-p)$. Moreover we decouple the four fermion
interaction term
in the above Hamiltonian by the Hatree-Fock approximation. Finally
 the mean field Hamiltonian of
the spinless Fermions has the expression:

\begin{equation}
H=\frac{J'}{2}\sum_{i,p=1}^{2M+1}(c_{i,p}^+c_{i,p+1}+H.C.)+\sum
_{i,p}^{2M}(c_{i,p}^+c_{i+1,2M+1-p}(-1)^{p+1}+H.C.)
\end{equation}
If we introduce a  Fourier transformation for the site indices, we have:
\begin{equation}
H=a\sum_{k,p=1}^{2M+1}(c_{k,p}^+c_{k,p+1}+H.C.)+\frac{J}{2}\sum
_{k,p}^{2M}(i\gamma_{k}(-1)^pc_{k,p}^+c_{k+1,2M+1-p}+H.C.)
\end{equation}
where $a={J' \over 2}$ $\gamma_k=-J\sin(k)$. The Hamiltonian then
 can be written 
in a form $H=\sum_{k}C_k^+h(k)C_k$ with h(k):

$$
h(k)=\left(
\begin{array}{cccccccccccccc}
0  & a  & .. & .. & .. & .. & .. & .. & .. & .. & .. & .. & ..& i\gamma\\
a  & 0  & a  & 0  & .. & .. & .. & .. & .. & .. & .. & .. & -i\gamma& ..\\
0  & a  & 0  & a  & .. & .. & .. & .. & .. & .. & .. & i\gamma  & ..& ..\\
.. & .. & a  & 0  & a  & .. & .. & .. & .. & .. & -i\gamma & .. & ..& ..\\
.. & .. & .. & a  & .. & .. & .. & .. & .. & i\gamma  & .. & .. & ..& ..\\
.. & .. & .. & .. & .. & .. & .. & .. & .. & .. & .. & .. & ..& ..\\
.. & .. & .. & .. & .. & .. & .. & .. & .. & .. & .. & .. & ..& ..\\
.. & .. & .. & .. & -i\gamma & .. & .. & .. & .. & .. & a  & .. & ..& ..\\
.. & .. & .. & i\gamma  & .. & .. & .. & .. & .. & a  & 0  & a  & ..& ..\\
.. & .. & -i\gamma & .. & .. & .. & .. & .. & .. & .. & a  & 0  & a & ..\\
.. & i\gamma & .. & .. & .. & .. & .. & .. & .. & .. & .. & a  & 0 & a \\
-i\gamma & .. &... & .. & .. & .. & .. & .. & .. & .. & .. & .. & a & 0
\end{array} \right)
$$

The above matrix contents 2M eigenvalues for a given wave number k
corresponding to the 2M individual bands separated by gaps. It can be proved
straightforwardly that for a given wave number k , half of the eigenvalues 
are less than zero and other half of them are great than zero. 
Furthermore we can also prove that zero is not an eigenvalue
of the above matrix for any non zero J'.  We will 
prove  the  two statements in the Appendix.
 This result shows clearly that half 
of the energy bands of the spinless Fermions is below zero energy and another
half is above zero. The energy gap between them is nonzero because zero is
not the eigenvalue of the above matrix for nonzero J' and arbitrary k.
Assuming there are no self-magnetization in one dimensional system, only the 
lower half
of the states is occupied by spinless Fermions in ground state. Therefor such 
spinless Fermion 
system is very similar with the traditional insulator in which the valence
band is fully occupied and the conductive band is fully empty in ground 
state. So a nonzero minimum energy is needed to excite the system from the 
ground state which indicates a spin excitation
energy gap.  For spin ladders with odd
number legs there exist odd numbers of energy band. Since only half of the
states is occupied in the ground state there must exist at least one
band which is partially occupied. This picture is very similar with the
traditional conductor in which exists at least one partial occupied band
(conduction band). Then the low energy excitations are gapless for
 odd-number-leg ladders.

Moreover, we calculate the gap size of the 2,4,6,8,and 10-leg
 spin ladder, the results
are shown in Fig.2 together with the experimental value. Although
our approach is quite rough, the result is in good agreement with the 
experimental value\cite{exp1,neu,sus,4leg}. 

The spin excitation spectra can also be obtained from the above mean field
approach, the spin wave dispersion is $\sqrt{({J' \over 2})^2+J^2\sin{k}^2}$ 
for the two-leg case.
It has two energy minimum, one is at 0 and another at $\pi$.
The shape of the spectra is not consistent with the numerical result which
has the spectra minimum located only at wave number $\pi$ and the maximum
is near $0.356\pi$. This is due to that the treatment for the phase term of 
eq.(3 )
is too rough. In the next section we will propose a more careful 
treatment of the
phase factor in eq.(3). We can then obtain a much more improved
 spin excitation
 spectra, which is
very close to the numerical results.

\vspace{1cm}
\centerline{\bf II. The mean field theory of 2-leg ladder}
\vspace{.5cm}

For the two-leg case, we can introduce two bipartite lattice labeled
as $\alpha$ and $\beta$. Following the same procedure shown in
the above section, the Hamiltonian for spinless Fermions becomes:

$$
H=J'\sum_i({1 \over 2}-\alpha_i^+\alpha_i)({1\over 2}-\beta_i^+\beta_i)+
  J\sum_i({1 \over 2}-\alpha_i^+\alpha_i)({1\over 2}-\beta_{i+1}^+\beta_{i+1})+
  J\sum_i({1 \over 2}-\alpha_{i+1}^+\alpha_{i+1})({1\over 2}-\beta_i^+\beta_i)+
$$
\begin{equation}
  {J' \over 2}\sum_i(\alpha_i^+\beta_i+h.c.)+
  {J  \over 2}\sum_i(\alpha_i^+\beta_{i+1}e^
  {i\pi(\beta_i^+\beta_i+\alpha_{i+1}^+\alpha_{i+1})}+h.c.)
  +{J \over 2}\sum_i(\beta_i^+\alpha_{i+1}+h.c.)
\end{equation}
In our mean field approach, different with the simple treatment used in the
previous section, we first replace the phase factor in (6) by it's 
average value:

$$
<e^{i\pi(\beta_i^+\beta_i+\alpha_{i+1}^+\alpha_{i+1})}>=
<(1-2\beta_i^+\beta_i)(1-2\alpha_{i+1}^+\alpha_{i+1})>
$$
$$
=-4\chi_1\chi_2
$$
where we define:
$$
\chi_1=<\beta_i^+\alpha_{i+1}>~~~~~~~~~~ \chi_2=<\alpha_{i+1}^+\beta_i>~~~~~~~~~
\chi_0=<\beta_i^+\alpha_i>
$$
Then the Fermion-Fermion interacting term $(1/2-\alpha_i^+\alpha_i)(1/2-
\beta_{i+1}^+\beta_{i+1})$ can be factorized as:
$$
(1/2-\alpha_i^+\alpha_i)(1/2-\beta_{i+1}^+\beta_{i+1})=
{1 \over 4}-\chi_0\beta_i^+\alpha_i-\chi_0^+\alpha_i^+\beta_i+\chi_0^+\chi_0
$$
We decouple the other two interacting term in the same manner and obtain
the following mean field Hamiltonian of spinless Fermions.

\begin{equation}
H_{MF}=\sum_k \gamma_k\alpha_k^+\beta_k +h.c.
\end{equation}
where:
$$
\gamma_k=\left(({J' \over 2}-J'\chi_0)+({J \over 2}-2J|\chi_2|^2-J\chi_1
-J\chi_2)\cos(k)\right)+
$$
$$
i\sin(k)\left(J\chi_2-J\chi_1-2J|\chi_2|^2-{J \over 2}
\right)
$$
Then the above Hamiltonian can be diagonalizyed as:
\begin{equation}
H_{MF}=\sum_k E_k\left( \tilde{\alpha}_k^+\tilde{\alpha}_k-\tilde{\beta}_k^+
\tilde{\beta}_k\right)
\end{equation}
in which :
$$
E_k=~~~~~~~~~~~~~~~~~~~~~~~~~~~~~~~~~~~~~~~~~~~~~~~~~~~~~~~~~~~~~~~~~~~~~~~~~~~~~~~~~~~~~~~~~~~~~~~~~~~~
$$
$$
\sqrt{\left(({J' \over 2}-J'\chi_0)+({J \over 2}-2J|\chi_2|^2-J\chi_1
-J\chi_2)\cos(k)\right)^2+\sin^2(k)\left(J\chi_2-J\chi_1-2J|\chi_2|^2-{J \over 2}
\right)^2}
$$  
The three parameters $\chi_1$,$\chi_2$ and $\chi_0$ is determined
selfconsistantly. The gap size obtained within the present approach
is shown in Fig.3 compared with the numerical results. Our results 
fits quite well to the numerical results in the parameter
 regime $J'/J>0.5$. For
the case of $J'=J$, the three parameters are found to be $\chi_1=-0.188J$
 $\chi_2=0.237J$ and $\chi_0=0.3867J$, and the gap is found to be 0.46J 
which is very close to the experimental value $(0.47\pm0.2)J$\cite{neu}.
In strong coupling limit $(J'>>J)$ our result fit well with the result obtained
by strong coupling expansion\cite{rei}, which shows 
$\Delta/J'\rightarrow 1$ when $J'/J\rightarrow \infty$.
But in the regime $J'/J<0.5$
our results deviated from the numerical results, and a nonzero gap persists
even at the case J'=0. So our mean field approach is valid only in the
intermidate and strong coupling regime. In the weak coupling regime our
mean field picture breaks due to the over estimation of the inter-chain
interaction. Since the phase factor in the hopping term in equation(6) is
replaced by its average value, it makes the hopping term within 
one chain being strongly modified by the motion of spinless Fermions in the
other chain, which is not valid for the weak coupling regime. 
We believe this approach is valid when the inter-chain coupling
is in the order of unit, but is not valid for the weak coupling case.

Another advantage of the present mean field approach is that in the case
of $J'=J$ it gives
the same spin wave dispersion predicted 
by the numerical calculation
as shown in Fig.4. The minimum of the spectra is at the wave number $\pi$,
and the maximum is at the wave number $0.356\pi$.  This result is in
good agreement with the numerical
result which has a minimum at $\pi$ and maximum at $0.3\pi$. We can also
calculate the two-magnon continuum from our mean field theory. 
The two magnon continuum
is proportional to 
$$
\int dt<S^z(q,t)S^z(-q,0)>e^{-i\omega t}
$$
which can
be transformed into the density-density correlation of 
the spinless Fermions by a
Jordan-Wigner transformation.
Then the two magnon excitation can be viewed as particle-hole excitation of
the spinless Fermions. The spectra of the two magnon excitation with several
specific q number is shown in Fig.5.
And the bottom(top) of the two magnon continuum is just 
the minimum(maximum) energy of hole-
particle excitation of the spinless Fermions for a given wave number. The result
is shown by dashed line(bottom) and dotted line(top) in Fig.4 which  
 fits  the numerical result quite well.\cite{num1,num2}.Compared to
the numerical method such as DMRG\cite{azz},quantum Monte Carlo 
and Lanczos method\cite{num1,num2},
our mean field theory based on the Jordan-Wigner Transformation gives
a more transparent understanding of the gap formation in even-number-leg
spin ladders and the low energy spin excitation.

The spin susceptibility
is also obtained by introduce a magnetic field in the original spinless
Fermion Hamiltonian, this term acts like the chemical potential:
$$
H=\sum_k \gamma_k\alpha_k^+\beta_k +h.c.-{1 \over N}\sum_{k}(\alpha_k^+
\alpha_k+\beta_k^+\beta_k)h
$$
The magnetization m then has the expression as:
$$
m={1 \over N}\sum_i(1-<\alpha_i^+\alpha_i>-<\beta_i^+\beta_i>)
$$
And the
spin susceptibility could be derived as:
$$
\chi_s=\frac{\partial m}{\partial h}
$$
We calculate the spin susceptibility in the case of $J'=J$ in a wide
range of temperature, the result is shown in Fig.6. Our result  
explains the temperature behavior of the spin susceptibility quite well,
and it is again in good agreement with the numerical results which gives a
maximum at $T=0.8J$.

\bigskip
\centerline{\bf IV. The Concluding Remarks}
\vspace{.5cm} 

In this paper we propose a mean field approach for spin ladders based on the
Jordan-Wigner transformation along an elaborately chosen
 path defined in this paper.
We show that in the mean field level that spin
gap is opened only in the even-number-leg-ladders and vanishes in the odd-
number-leg-ladders. It gives a very simple picture of the formation
and vanishing of the spin gap in the above mentioned two type of spin ladders.
The spin ladders with even number legs formed a 'insulator' like band for
spinless Fermions, whereas in odd-number-leg ladders the band structure of
the spinless Fermions is 'metal' like.

Then we take a more careful study of the 2-leg ladder. Particular for the
$J=J'$ case the magnitude of the gap found in
our approach is in good agreement with both the numerical result and the 
experimental result. Further the spin excitation spectra and the uniform
 susceptibility are also calculated based on our mean field treatment.
The dispersion relation of the spin excitation spectra obtained by our 
mean field theory is very similar with the numerical result, which
has its maximum locating between $0$ and $\pi$. The uniform susceptibility
is 
 also consistent with the numerical results, which predicts a maximum
at $T=0.8J$.

\newpage
\bigskip
\centerline{\bf Appendix}

In the Appendix we prove that matrix h(k) in section II has the two following
properties:(i)
If $\lambda$ is a eigenvalue of the matrix, $-\lambda$ is also a eigenvalue
 of it.
(ii) Zero can not be a eigenvalue of the matrix with any nonzero J'. 

First we divide the Hermite matrix h(k) into its real and imaginary part
 $h(k)=
A+iB$, in which the matrix A,B read as:

$$
A_{ij}=a\delta_{i,j+1}+a\delta_{i,j-1}~~~~~~~~~~~ B_{ij}=(-1)^{i+1}\delta_{i,2M+1-j}\gamma
$$
One can easily find that the matrix A and B satisfies some relations:
$$
(I)~~~~~~~~~~~~~~~~ K\cdot h(k) \cdot K=-h(k)~~~~~~~~~~~~~~~~~~~~~~~~~~~~~~~~~~~~~~~~~~~~~~~~~~~~~~~~~~~
$$

$$
(II)~~~~~~~~~~~~~~~A^T=A~~~~B^T=-B~~~~B^{-1}=\gamma^{-2}B~~~~B^{-1}AB=-A~~~~~~~~~~~~~~~~~~~
$$
in which $K_{ij}=\delta_{ij}(-1)^i$ and $K_{il}K_{lj}=\delta_{ij}$

Based on the above equations we can prove the two properties straightforwardly.
For (i)if we have $h(k)x=\lambda x$, we can multiply the matrix K to the both side of
the above equation: $K\cdot h(k)\cdot K\cdot Kx=\lambda Kx$, then we have 
$h(k)(Kx)=-\lambda (Kx)$.

For property (ii), we can prove it as follows. First if B=0 the conclusion is 
obviously true because the determinant of matrix A is nonzero for $J'\neq 0$
and this makes the equation $Ax=0$ can not be satisfied unless x=0. Next when
$B\neq 0$ we have :
$$
(A+iB)x=0~~or~~~~~equivalently~~~~~B^{-1}Ax=-ix
$$
For matrix $C=B^{-1}A$, we have
$$
C^+=[B^{-1}A]^{+}=\gamma^{-2}A^+B^+=-\gamma^{-2}AB=-(B^{-1})(\gamma^{-2}B)AB
=(B^{-1})A=C
$$
Then the matrix C is Hermite, and it can not has a imaginary eigenvalue, so
the equation $(A+iB)\cdot x=0$ can not be satisfied for nonzero x.

In the above proof we used the relation (II). Then from the above paragraph 
we prove
the two properties used in section II.

\newpage
\bigskip
\centerline{\bf FIGURE CAPTIONS}
\vspace{1.0ex}
\parskip=2pt
\baselineskip=14pt

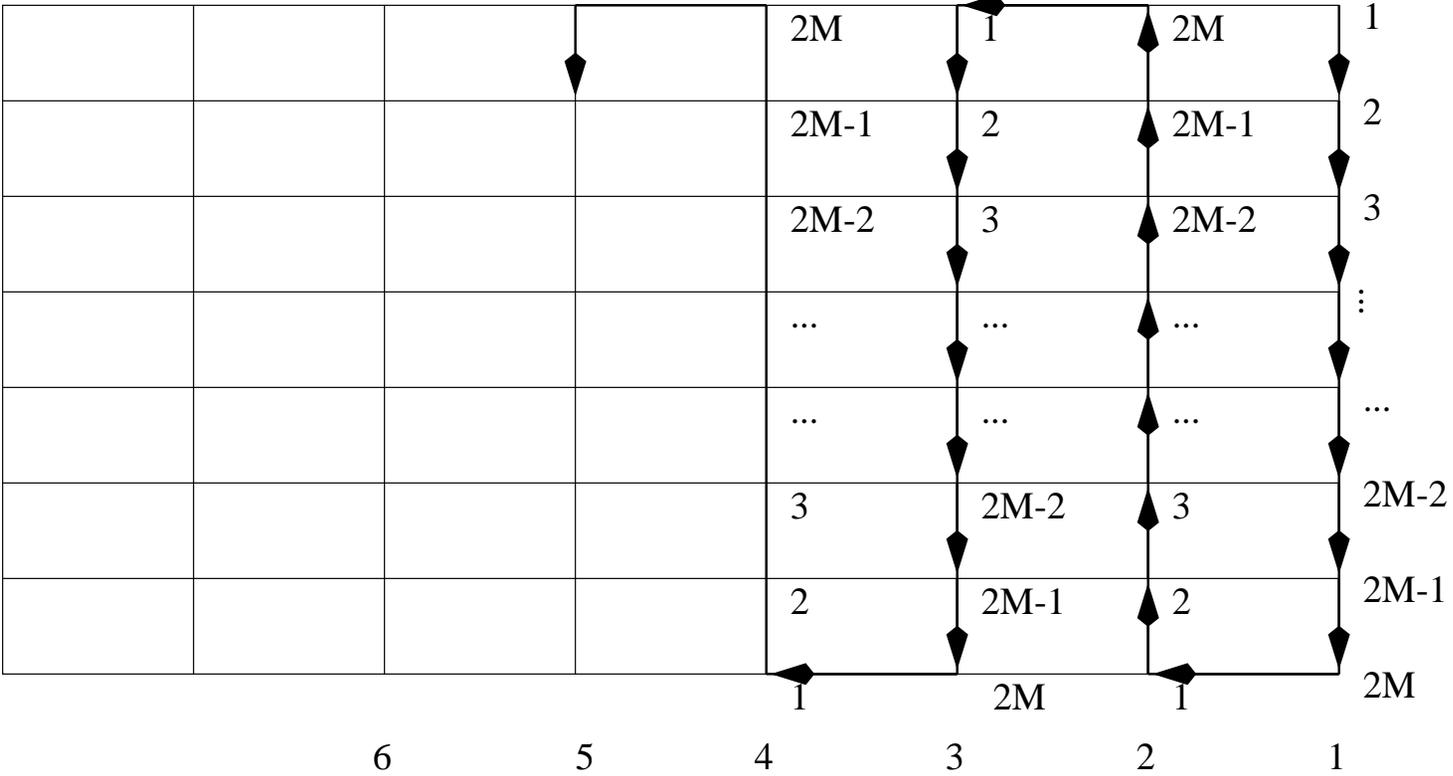
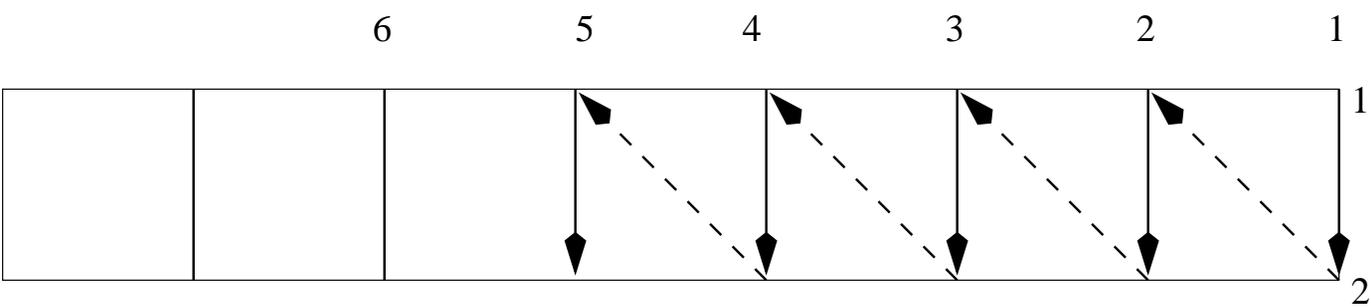
Fig.~1: To perform the Jordan-Wigner transformation the sites are put in a
particular queue in our present study(a) and in paper \cite{azz}(b).

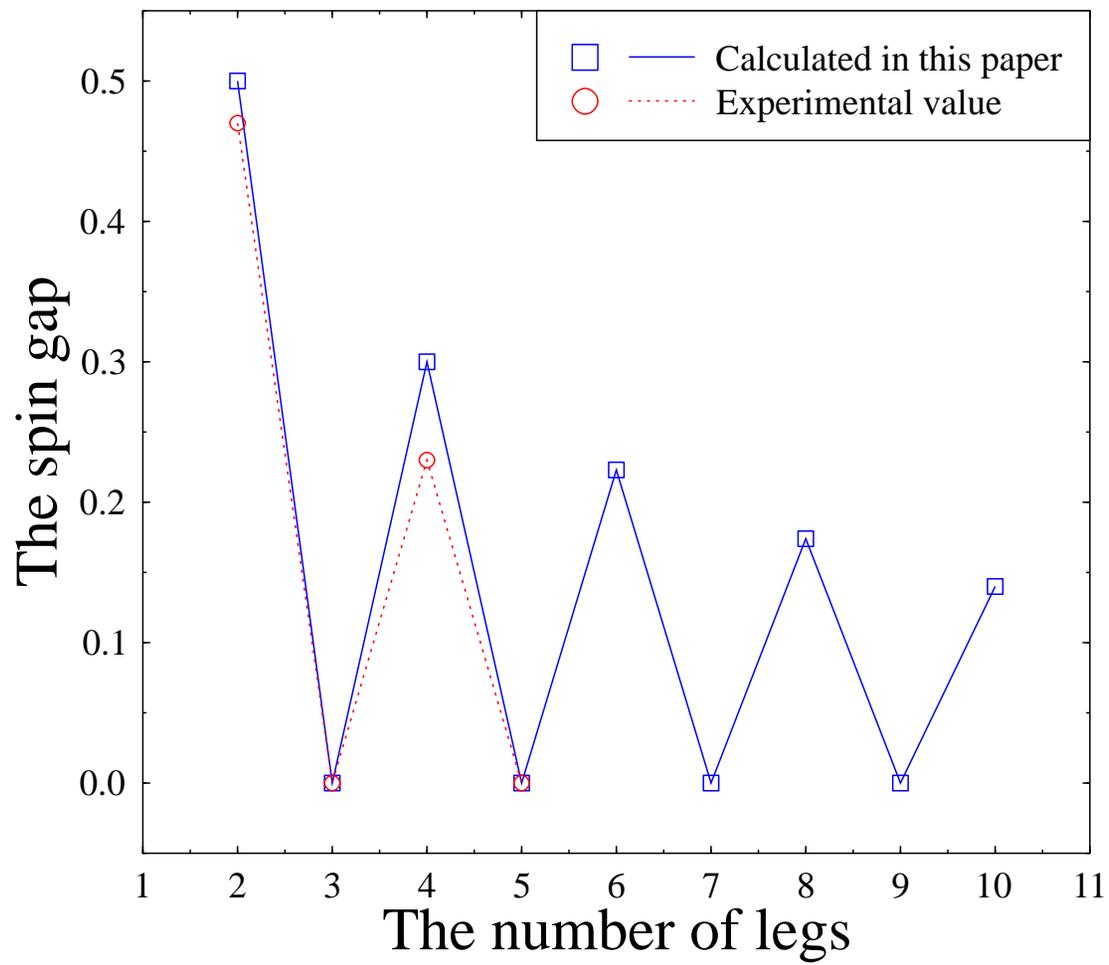
Fig.~2: The spin gap for n-leg spin ladders calculated in our mean 
field approach which is compared by the experimental value.\cite{neu,4leg}.

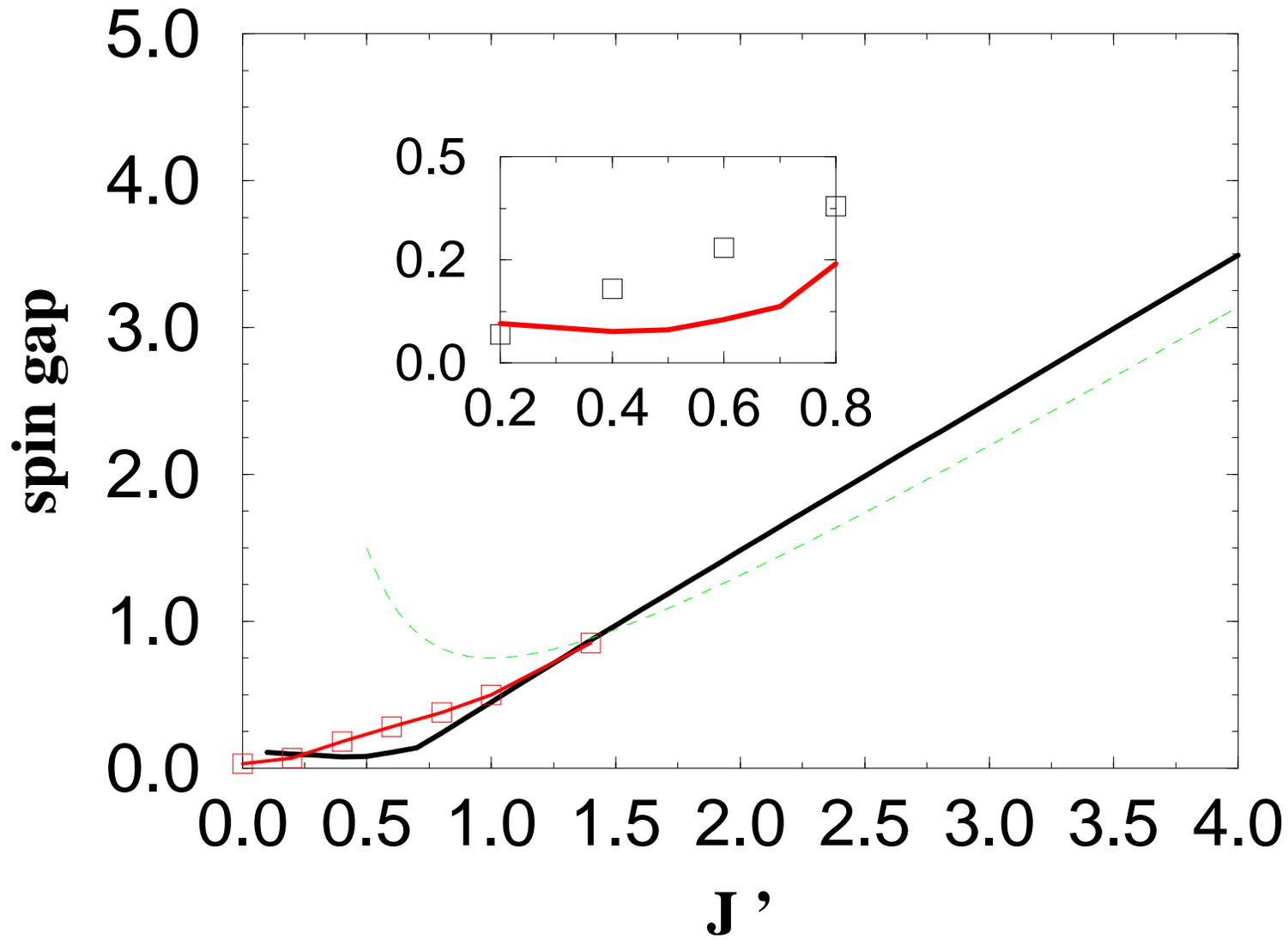
Fig.~3: The solid line shows the spin gap obtained by the self consistent
procedure for 2-leg  ladder as a function of inter chain coupling $J'$.The
dashed line is the result of strong coupling expantion\cite{rei}.
The squares shows the numerical result of paper\cite{num1}. The inset shows
the results with J'/J less than 0.8. 

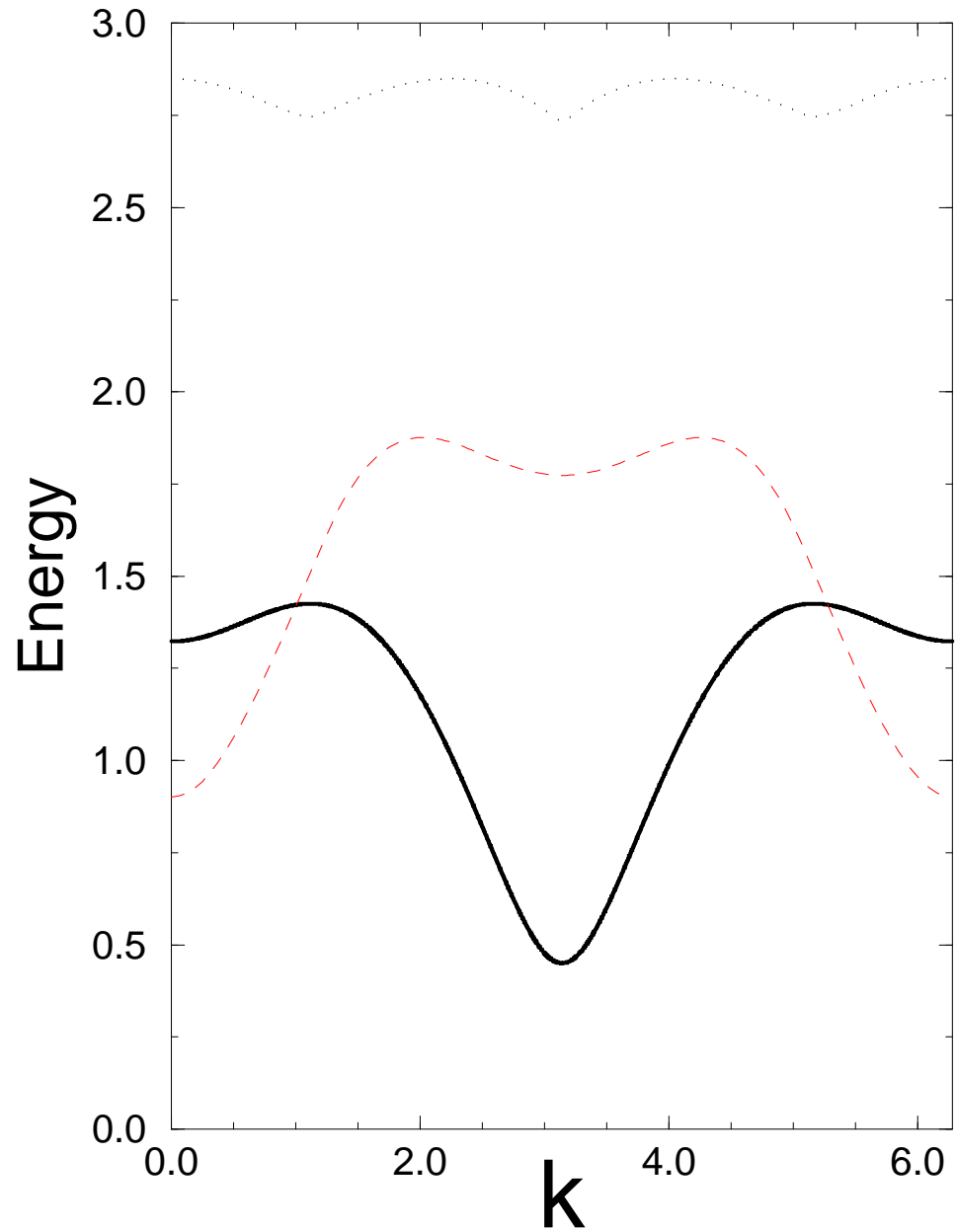
Fig.~4: The solid line is the dispersion of the spinless Fermions for 2-leg 
ladder calculated in our mean field approach.The dashed(dotted) line is the
bottom(top) of the two magnon continuum.

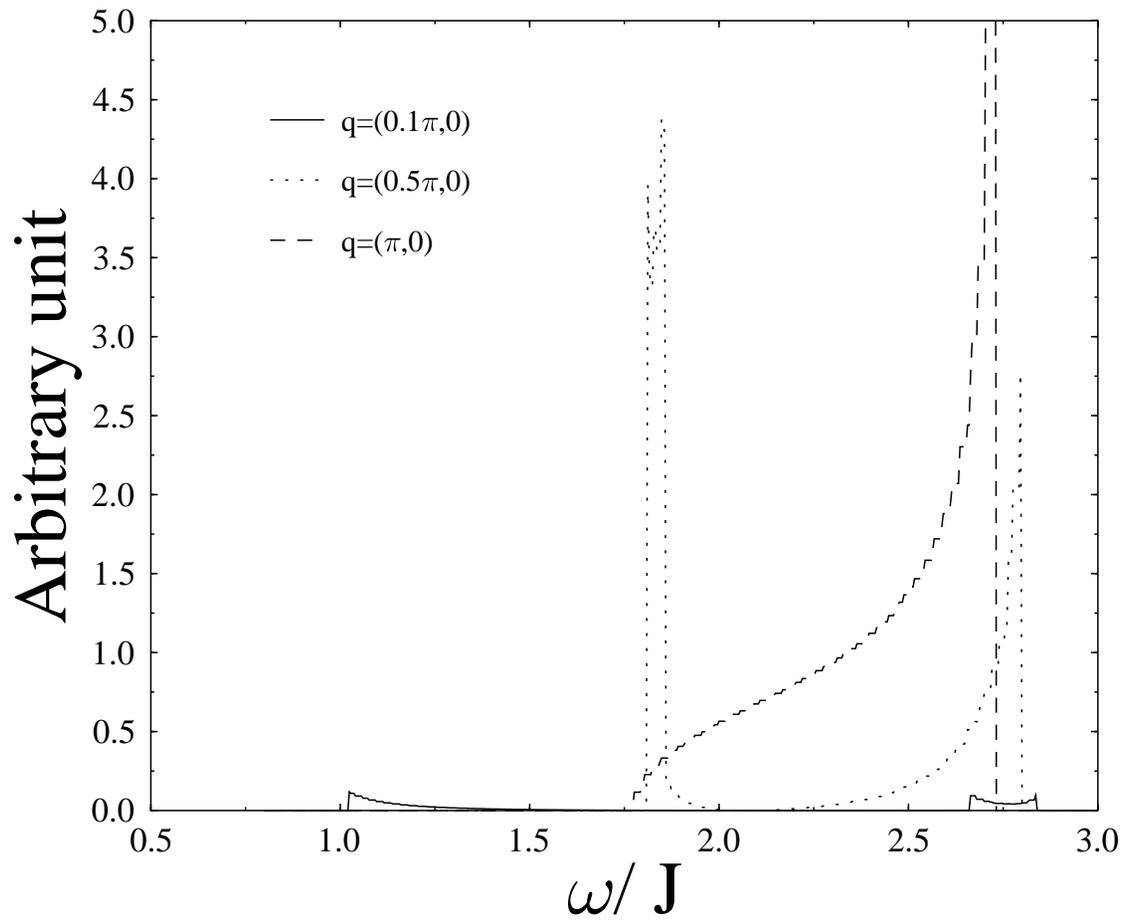
Fig.~5: The spectra of two-magnon excitation with $q=0.1\pi$(dotted line),
$0.5\pi$(solid line),$\pi$(dashed line).  

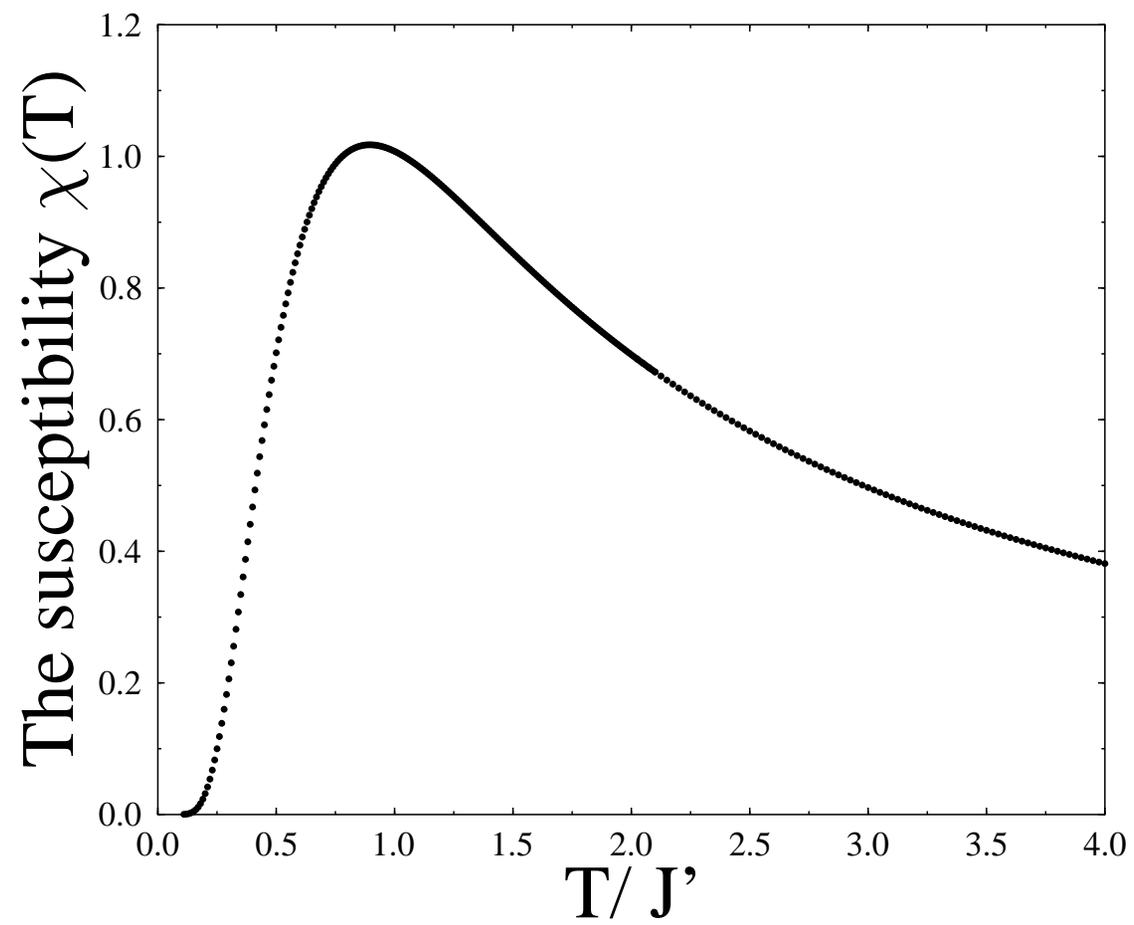
Fig.~6: The spin susceptibility of 2-leg ladder in the case of $J=J'$.

\end{document}